\documentclass[english,aps,manuscript]{revtex4}
\usepackage[T1]{fontenc}
\usepackage[latin1]{inputenc}
\usepackage{amssymb}

\makeatletter

\newcommand{\ben}{\begin{enumerate}}
\newcommand{\een}{\end{enumerate}}
\newcommand{\be}{\begin{equation}}
\newcommand{\ee}{\end{equation}}
\newcommand{\bse}{\begin{subequation}}
\newcommand{\ese}{\end{subequation}}
\newcommand{\bea}{\begin{eqnarray}}
\newcommand{\eea}{\end{eqnarray}}
\newcommand{\bc}{\begin{center}}
\newcommand{\ec}{\end{center}}

\usepackage{doublespace}
\usepackage{epsfig}

\usepackage{babel}
\makeatother
\begin{document}

\title{Wigner Function Formalism for Zero Magnetic Field Spin Dependent
Resonant Tunneling Structures }

\author{Mehmet Burcin Unlu, Hong-Liang Cui}

\address{Department of Physics and Engineering Physics, Stevens Institute
of Technology, Hoboken, NJ 07030, USA}

\begin{abstract}
We develop a Wigner function representation of the quantum transport
theory of the conduction band electrons in Rashba effect resonant
tunneling structures with a phonon bath. In narrow band gap heterostructures,
spin splitting occurs mainly as a result of inversion asymmetry in
the spatial dependence of the potential or as a result of external
electric field. This \char`\"{}zero magnetic field spin splitting\char`\"{}
is due to the Rashba term in the effective mass Hamiltonian. The quantum
transport equations are derived using multi-band non-equilibrium Green's
function formulation in generalized Kadanoff-Baym ansatz. 
\end{abstract}

\keywords{Rashba effect, resonant tunneling structure, Wigner function, spin
transport.}

\pacs{71.70.Ej, 72.25.Dc, 73.63.Hs }

\maketitle
Zero magnetic field spin splitting \cite{Winkler2} of the conduction
band states in bulk and compound semiconductors has attracted great
interest recently. The possibility of a spin polarized current source
based on conventional nonmagnetic semiconductors has led many researchers
to investigate the possibility of obtaining efficient spin-polarized
current source, a necessity in spintronics. Bulk inversion asymmetry
(as in zincblende semiconductors) dominates in large band gap semiconductors
whereas the structural inversion asymmetry ( the Rashba effect ) becomes
important in narrow band gap semiconductors. 

The Rashba effect predicts a spin splitting of the conduction band
linear in $k_{\parallel}$. For example, the spin splitting in the
lowest conduction subband exceeds $0.02$ $eV$ for in-plane wave
vector $k_{\parallel}=0.05$ $\textrm{Å}^{-1}$ \cite{Ting}. A similar
splitting occurs in the light hole band proportional to $k_{\parallel}$
and in the heavy hole band proportional to $k_{\parallel}^{3}$ \cite{Winkler}.
The case of conduction band electrons is more attractive for the spin
transport devices since these have a much longer spin relaxation time.

Wigner function modeling of charge transport in multi-barrier resonant
tunneling structures has been quite popular in the literature due
to its success in dealing with the dissipation and the open boundary
conditions \cite{Buot}. Similarly, it is expected that one should
be able to model the spin transport in the tunneling structures using
the Wigner function. 

We derive two conduction spin subband Wigner function equations for
zero magnetic field resonant tunneling structures. These equations
are derived for the first time in the literature. When the spin-orbit
interaction is included in the single conduction band effective mass
Hamiltonian, the Wigner function becomes a $2\times2$ matrix in spin
subspace. The diagonal elements $f_{\uparrow\uparrow}$ and $f_{\downarrow\downarrow}$
are the spin up, spin down electron distribution functions respectively.
The off-diagonal elements $f_{\uparrow\downarrow}$ and $f_{\downarrow\uparrow}$
describe the spin coherence. Note that $f_{\downarrow\uparrow}=f_{\uparrow\downarrow}^{*}$. 

The Rashba effect resonant tunneling structures have been discussed
from both theoretical and experimental points of view in the literature.
It was shown that an asymmetric double barrier resonant tunneling
diode (DB RTD) might provide spin polarization above 50 \% \cite{Erasmos}.
Koga et al proposed a triple barrier resonant tunneling diode (TB
RTD) that can reveal a high degree of polarization 99.9 \% \cite{Koga}.
Recently Ting and Cartoixa proposed the resonant inter-band tunneling
spin filter \cite{David}.

If the $z$- axis is chosen as the growth direction, the effective
mass Hamiltonian with the spin-orbit term can be written as

\[
H=\hat{p}_{z}\frac{1}{2m^{*}(z)}\hat{p}_{z}+\frac{{\bf p}_{\parallel}^{2}}{2m^{*}(z)}+\frac{\alpha}{\hbar}({\bf \sigma}\times{\bf p})_{z}\]
where $\sigma$ denotes the Pauli spin matrices, ${\bf p}$ is the
momentum operator, ${\bf p}_{\parallel}=(p_{x},p_{y},0)$ is the in-plane
momentum vector. We assume that the in-plane momentum, ${\bf p}_{\parallel}$,
is conserved across the device. $\alpha$ is the spin-orbit coupling
constant (Rashba constant) and it is material dependent (inversely
proportional to the energy gap and the effective mass) and proportional
to the interface electric field along the growth direction. Therefore
it is possible to tune this parameter by applying an external bias.
The experimental value of the Rashba constant is in the order of $10^{-12}eV.m$. 

The Hamiltonian for Rashba effect resonant tunneling structure can
be written as, 

\begin{equation}
H=\left[\begin{array}{cc}
H_{\uparrow\uparrow} & H_{\uparrow\downarrow}\\
H_{\downarrow\uparrow} & H_{\downarrow\downarrow}\end{array}\right]=\left[\begin{array}{cc}
\hat{p}_{z}\frac{1}{2m^{*}(z)}\hat{p}_{z}+\frac{p_{\parallel}^{2}}{2m^{*}(z)}+E_{c}(z) & i\frac{\alpha}{\hbar}p_{-}\\
-i\frac{\alpha}{\hbar}p_{+} & \hat{p}_{z}\frac{1}{2m^{*}(z)}\hat{p}_{z}+\frac{p_{\parallel}^{2}}{2m^{*}(z)}+E_{c}(z)\end{array}\right]\end{equation}
 where $\hat{p}_{z}=-i\hbar\frac{\partial}{\partial z}$ , $p_{\pm}=p_{x}\pm ip_{y}$
, and $E_{c}(z)=E_{c}+V(z)$. $E_{c}$ is the conduction band edge
and $V(z)$ is the self-consistent potential. 

Throughout the paper, $\sigma$ denotes the spin index and there is
a summation over repeated indices. The four-dimensional, (3+1), crystal
momentum and its conjugate variable lattice coordinate are represented
as $p=({\bf p},E)$, $r=({\bf R},T)$. Under the assumption that the
self-energies are slowly varying with respect to the center of mass
coordinates, generalized Kadanoff-Baym equation for in phase-space-energy-time
domain can be written as \cite{Unlu},

\begin{eqnarray}
i\hbar\frac{\partial}{\partial T}G_{\sigma\sigma^{'}}^{<}(p,r) & = & \frac{1}{(h^{4})^{2}}\int dp_{2}dr_{2}K_{H_{\sigma\beta}}^{c}(p,r-r_{2};r,p-p_{2})G_{\beta\sigma^{'}}^{<}(p_{2},r_{2})\nonumber \\
 &  & +\Sigma_{\sigma\beta}^{>}(p,r)G_{\beta\sigma^{'}}^{<}(p,r)-\Sigma_{\sigma\beta}^{<}(p,r)G_{\beta\sigma^{'}}^{>}(p,r)\label{GKB}\end{eqnarray}
where 

\[
K_{A}^{c}(p,r-r_{2};r,p-p_{2})=K_{A}^{+}(p,r-r_{2};r,p-p_{2})-K_{A}^{-}(p,r-r_{2};r,p-p_{2})\]
 and 

\[
K_{A}^{\pm}(p,r-r_{2};r,p-p_{2})=\int dp_{1}dr_{1}\exp(\frac{i}{\hbar}p_{1}.(r-r_{2}))\exp(\frac{i}{\hbar}r_{1}.(p-p_{2}))A(p\pm\frac{p_{1}}{2},r\mp\frac{r_{1}}{2}).\]

The self-energy function for the electron-phonon scattering in phase-space-energy-time
representation can be written as

\begin{equation}
\Sigma_{\sigma\sigma^{'}}^{>,<}(p,r)=\frac{i}{h^{4}}\int dqG_{\sigma\sigma^{'}}^{>,<}(p+q)D^{>,<}(q).\end{equation}
Assuming the phonon bath is in equilibrium, the Fourier transforms
of the phonon Green's functions can be written as,

\begin{equation}
D^{<}(\mathbf{q},E^{'})=-ihM_{\mathbf{q}}^{2}[(N_{\mathbf{q}}+1)\delta(E^{'}-\Omega_{\mathbf{q}})+N_{\mathbf{q}}\delta(E^{'}+\Omega_{\mathbf{q}})],\end{equation}

\begin{equation}
D^{>}(\mathbf{q},E^{'})=-ihM_{\mathbf{q}}^{2}[(N_{\mathbf{q}}+1)\delta(E^{'}+\Omega_{\mathbf{q}})+N_{\mathbf{q}}\delta(E^{'}-\Omega_{\mathbf{q}})]\end{equation}
where $M_{\mathbf{q}}$ is the electron-phonon scattering matrix element.
Therefore, inclusion of the phonon scattering gives the following
scattering functions

\begin{equation}
\Sigma_{\sigma\sigma^{'}}^{<}=\sum_{\eta=+1,-1}\frac{1}{h^{3}}\int d{\bf q}G_{\sigma\sigma^{'}}^{<}(\mathbf{p}+\mathbf{q},E+\eta\Omega_{\mathbf{q}},r)M_{\mathbf{q}}^{2}(N_{\mathbf{q}}+\frac{1}{2}+\frac{1}{2}\eta),\label{eq:1}\end{equation}

\begin{equation}
\Sigma_{\sigma\sigma^{'}}^{>}=\sum_{\eta=+1,-1}\frac{1}{h^{3}}\int d{\bf q}G_{\sigma\sigma^{'}}^{>}(\mathbf{p}+\mathbf{q},E+\eta\Omega_{\mathbf{q}},r)M_{\mathbf{q}}^{2}(N_{{\bf \mathbf{q}}}+\frac{1}{2}-\frac{1}{2}\eta).\label{eq:2}\end{equation}
The free generalized Kadanoff-Baym (FGKB) \cite{Langreth}, \cite{Koch}
ansatz for a two spin band system can be stated as,

\begin{equation}
-iG_{\sigma\sigma^{'}}^{<}(\mathbf{p},E,\mathbf{r},T)=hf_{\sigma\sigma^{'}}(\mathbf{p},\mathbf{r},T)\delta(E-\frac{E_{\sigma}(\mathbf{p})+E_{\sigma^{'}}({\bf \mathbf{p}})}{2})\label{eq:3}\end{equation}

\begin{equation}
iG_{\sigma\sigma^{'}}^{>}(\mathbf{p},E,\mathbf{r},T)=-h(\delta_{\sigma\sigma^{'}}-f_{\sigma\sigma^{'}}(\mathbf{p},\mathbf{r},T))\delta(E-\frac{E_{\sigma}(\mathbf{p})+E_{\sigma^{'}}(\mathbf{p})}{2})\label{eq:4}\end{equation}
The spin subband Wigner function is found by taking the energy integral
of the $G_{\sigma\sigma^{'}}^{<}$:

\begin{equation}
f_{\sigma\sigma^{'}}(\mathbf{p},\mathbf{R},T)=\int dE(-i)G_{\sigma\sigma^{'}}^{<}(\mathbf{p},E;\mathbf{R},T).\label{wigner1}\end{equation}

Substituting the equations (\ref{eq:1}), (\ref{eq:2}) to equation
(\ref{GKB}) together with the FGKB ansatz (\ref{eq:3}), (\ref{eq:4})
and using the expression (\ref{wigner1}) gives the desired set of
equations. As a result, the spin up Wigner function for a Rashba effect
RTD in weakly contact with a phonon heat bath can be written as

\begin{eqnarray}
\hbar\frac{\partial f_{\uparrow\uparrow}(p_{\parallel},p_{z},z,t)}{\partial T} & = & \frac{1}{2h}\int dz_{1}dp_{z2}sin[\frac{1}{\hbar}z_{1}(p_{z}-p_{z2})](\frac{1}{m^{*}(z-\frac{z_{1}}{2})}-\frac{1}{m^{*}(z+\frac{z_{1}}{2})})p_{z}^{2}f_{\uparrow\uparrow}(p_{\parallel},p_{z2},z,T)\nonumber \\
 &  & -\frac{1}{8\pi}\int dz_{1}dp_{z2}cos[\frac{1}{\hbar}z_{1}(p_{z}-p_{z2})](\frac{1}{m^{*}(z-\frac{z_{1}}{2})}+\frac{1}{m^{*}(z+\frac{z_{1}}{2})})p_{z}\frac{\partial}{\partial z}f_{\uparrow\uparrow}(p_{\parallel},p_{z2},z,T)\nonumber \\
 &  & -\frac{1}{8\pi}\int dz_{1}dp_{z2}cos[\frac{1}{\hbar}z_{1}(p_{z}-p_{z2})]\frac{\partial}{\partial z}(\frac{1}{m^{*}(z-\frac{z_{1}}{2})}+\frac{1}{m^{*}(z+\frac{z_{1}}{2})})p_{z}f_{\uparrow\uparrow}(p_{\parallel},p_{z2},z,T)\nonumber \\
 &  & -\frac{1}{8h}\int dz_{1}dp_{z2}sin[\frac{1}{\hbar}z_{1}(p_{z}-p_{z2})]\frac{\partial}{\partial z}(\frac{1}{m^{*}(z-\frac{z_{1}}{2})}-\frac{1}{m^{*}(z+\frac{z_{1}}{2})})\frac{\partial}{\partial z}f_{\uparrow\uparrow}(p_{\parallel},p_{z2},z,T)\nonumber \\
 &  & +\frac{1}{2h}\int dz_{1}dp_{z2}sin[\frac{1}{\hbar}z_{1}(p_{z}-p_{z2})](\frac{1}{m^{*}(z-\frac{z_{1}}{2})}-\frac{1}{m^{*}(z+\frac{z_{1}}{2})})p_{\parallel}^{2}f_{\uparrow\uparrow}(p_{\parallel},p_{z2},z,T)\nonumber \\
 &  & +\frac{1}{h}\int dz_{1}dp_{z2}sin(\frac{1}{\hbar}z_{1}(p_{z}-p_{z2}))[E_{c}(z-\frac{z_{1}}{2})-E_{c}(z+\frac{z_{1}}{2})]f_{\uparrow\uparrow}(p_{\parallel},p_{z2},z,T)\nonumber \\
 &  & +\frac{\alpha}{\hbar}p_{x}(f_{\uparrow\downarrow}+f_{\downarrow\uparrow})+i\frac{\alpha}{\hbar}p_{y}(f_{\uparrow\downarrow}-f_{\downarrow\uparrow})\nonumber \\
 &  & +\sum_{\sigma=\uparrow,\downarrow}\sum_{\eta=+1,-1}\frac{1}{h^{3}}\int d\mathbf{q}(\delta_{\uparrow\sigma}-f_{\uparrow\sigma}(\mathbf{p}+\mathbf{q},\mathbf{r},T))f_{\sigma\uparrow}({\bf \mathbf{p}},\mathbf{r},T)\nonumber \\
 &  & \times\delta(\frac{E_{\uparrow}({\bf \mathbf{p}}+{\bf \mathbf{q}})+E_{\sigma}({\bf \mathbf{p}}+{\bf \mathbf{q}})}{2}-\frac{E_{\sigma}({\bf \mathbf{p}})+E_{\uparrow}(\mathbf{p})}{2}+\eta\Omega_{\mathbf{q}})M_{\mathbf{q}}^{2}(N_{\mathbf{q}}+\frac{1}{2}-\frac{1}{2}\eta)\nonumber \\
 &  & -\sum_{\sigma=\uparrow,\downarrow}\sum_{\eta=+1,-1}\frac{1}{h^{3}}\int d\mathbf{q}f_{\uparrow\sigma}(\mathbf{{\bf \mathbf{p}}+{\bf \mathbf{q}}},\mathbf{r},T)(\delta_{\sigma\uparrow}-f_{\sigma\uparrow}({\bf \mathbf{p}},\mathbf{r},T))\nonumber \\
 &  & \times\delta(\frac{E_{\uparrow}({\bf \mathbf{p}}+{\bf \mathbf{q}})+E_{\sigma}({\bf \mathbf{p}}+{\bf \mathbf{q}})}{2}-\frac{E_{\sigma}({\bf \mathbf{p}})+E_{\uparrow}(\mathbf{p})}{2}+\eta\Omega_{\mathbf{q}})M_{\mathbf{q}}^{2}(N_{\mathbf{q}}+\frac{1}{2}+\frac{1}{2}\eta).\end{eqnarray}
The first five terms at right hand side of the above equation correspond
to the drift terms. These terms are quite complicated comparing to
the drift term in the usual single band Wigner function equation since
the effective mass is taken to be position dependent. As a result,
the effective mass becomes nonlocal. The sixth term gives the potential
term and it is calculated self-consistently by coupling to Poisson
equation. The seventh term is the subband mixing term. The last two
terms correspond to the electron-phonon scattering. The relaxation
time approximation can be made for these. Similarly, the inter spin
band polarization of the spin coherence becomes 

\begin{eqnarray}
\hbar\frac{\partial f_{\uparrow\downarrow}(p_{\parallel},p_{z},z,t)}{\partial T} & = & \frac{1}{2h}\int dz_{1}dp_{z2}sin[\frac{1}{\hbar}z_{1}(p_{z}-p_{z2})](\frac{1}{m^{*}(z-\frac{z_{1}}{2})}-\frac{1}{m^{*}(z+\frac{z_{1}}{2})})p_{z}^{2}f_{\uparrow\downarrow}(p_{\parallel},p_{z2},z,T)\nonumber \\
 &  & -\frac{1}{8\pi}\int dz_{1}dp_{z2}cos[\frac{1}{\hbar}z_{1}(p_{z}-p_{z2})](\frac{1}{m^{*}(z-\frac{z_{1}}{2})}+\frac{1}{m^{*}(z+\frac{z_{1}}{2})})p_{z}\frac{\partial}{\partial z}f_{\uparrow\downarrow}(p_{\parallel},p_{z2},z,T)\nonumber \\
 &  & -\frac{1}{8\pi}\int dz_{1}dp_{z2}cos[\frac{1}{\hbar}z_{1}(p_{z}-p_{z2})]\frac{\partial}{\partial z}(\frac{1}{m^{*}(z-\frac{z_{1}}{2})}+\frac{1}{m^{*}(z+\frac{z_{1}}{2})})p_{z}f_{\uparrow\downarrow}(p_{\parallel},p_{z2},z,T)\nonumber \\
 &  & -\frac{1}{8h}\int dz_{1}dp_{z2}sin[\frac{1}{\hbar}z_{1}(p_{z}-p_{z2})]\frac{\partial}{\partial z}(\frac{1}{m^{*}(z-\frac{z_{1}}{2})}-\frac{1}{m^{*}(z+\frac{z_{1}}{2})})\frac{\partial}{\partial z}f_{\uparrow\downarrow}(p_{\parallel},p_{z2},z,T)\nonumber \\
 &  & +\frac{1}{2h}\int dz_{1}dp_{z2}sin[\frac{1}{\hbar}z_{1}(p_{z}-p_{z2})](\frac{1}{m^{*}(z-\frac{z_{1}}{2})}-\frac{1}{m^{*}(z+\frac{z_{1}}{2})})p_{\parallel}^{2}f_{\uparrow\downarrow}(p_{\parallel},p_{z2},z,T)\nonumber \\
 &  & +\frac{1}{h}\int dz_{1}dp_{z2}sin(\frac{1}{\hbar}z_{1}(p_{z}-p_{z2}))[E_{c}(z-\frac{z_{1}}{2})-E_{c}(z+\frac{z_{1}}{2})]f_{\uparrow\downarrow}(p_{\parallel},p_{z2},z,T)\\
 &  & +\frac{\alpha}{\hbar}p_{x}(f_{\downarrow\downarrow}-f_{\uparrow\uparrow})+i\frac{\alpha}{\hbar}p_{y}(f_{\uparrow\uparrow}-f_{\downarrow\downarrow})\nonumber \\
 &  & +\sum_{\sigma=\uparrow,\downarrow}\sum_{\eta=+1,-1}\frac{1}{h^{3}}\int d\mathbf{q}(\delta_{\uparrow\sigma}-f_{\uparrow\sigma}(\mathbf{p}+\mathbf{q},\mathbf{r},T))f_{\sigma\downarrow}({\bf \mathbf{p}},\mathbf{r},T)\nonumber \\
 &  & \times\delta(\frac{E_{\uparrow}({\bf \mathbf{p}}+{\bf \mathbf{q}})+E_{\sigma}({\bf \mathbf{p}}+{\bf \mathbf{q}})}{2}-\frac{E_{\sigma}({\bf \mathbf{p}})+E_{\downarrow}(\mathbf{p})}{2}+\eta\Omega_{\mathbf{q}})M_{\mathbf{q}}^{2}(N_{\mathbf{q}}+\frac{1}{2}-\frac{1}{2}\eta)\nonumber \\
 &  & -\sum_{\sigma=\uparrow,\downarrow}\sum_{\eta=+1,-1}\frac{1}{h^{3}}\int d\mathbf{q}f_{\uparrow\sigma}(\mathbf{{\bf \mathbf{p}}+{\bf \mathbf{q}}},\mathbf{r},T)(\delta_{\sigma\downarrow}-f_{\sigma\downarrow}({\bf \mathbf{p}},\mathbf{r},T))\nonumber \\
 &  & \times\delta(\frac{E_{\uparrow}({\bf \mathbf{p}}+{\bf \mathbf{q}})+E_{\sigma}({\bf \mathbf{p}}+{\bf \mathbf{q}})}{2}-\frac{E_{\sigma}({\bf \mathbf{p}})+E_{\downarrow}(\mathbf{p})}{2}+\eta\Omega_{\mathbf{q}})M_{\mathbf{q}}^{2}(N_{\mathbf{q}}+\frac{1}{2}+\frac{1}{2}\eta).\end{eqnarray}
 Note that $f_{\downarrow\uparrow}=f_{\uparrow\downarrow}^{*}$. Finally
the spin down component of Wigner function is given as, 

\begin{eqnarray}
\hbar\frac{\partial f_{\downarrow\downarrow}(p_{\parallel},p_{z},z,t)}{\partial T} & = & \frac{1}{2h}\int dz_{1}dp_{z2}sin[\frac{1}{\hbar}z_{1}(p_{z}-p_{z2})](\frac{1}{m^{*}(z-\frac{z_{1}}{2})}-\frac{1}{m^{*}(z+\frac{z_{1}}{2})})p_{z}^{2}f_{\downarrow\downarrow}(p_{\parallel},p_{z2},z,T)\nonumber \\
 &  & -\frac{1}{8\pi}\int dz_{1}dp_{z2}cos[\frac{1}{\hbar}z_{1}(p_{z}-p_{z2})](\frac{1}{m^{*}(z-\frac{z_{1}}{2})}+\frac{1}{m^{*}(z+\frac{z_{1}}{2})})p_{z}\frac{\partial}{\partial z}f_{\downarrow\downarrow}(p_{\parallel},p_{z2},z,T)\nonumber \\
 &  & -\frac{1}{8\pi}\int dz_{1}dp_{z2}cos[\frac{1}{\hbar}z_{1}(p_{z}-p_{z2})]\frac{\partial}{\partial z}(\frac{1}{m^{*}(z-\frac{z_{1}}{2})}+\frac{1}{m^{*}(z+\frac{z_{1}}{2})})p_{z}f_{\downarrow\downarrow}(p_{\parallel},p_{z2},z,T)\nonumber \\
 &  & -\frac{1}{8h}\int dz_{1}dp_{z2}sin[\frac{1}{\hbar}z_{1}(p_{z}-p_{z2})]\frac{\partial}{\partial z}(\frac{1}{m^{*}(z-\frac{z_{1}}{2})}-\frac{1}{m^{*}(z+\frac{z_{1}}{2})})\frac{\partial}{\partial z}f_{\downarrow\downarrow}(p_{\parallel},p_{z2},z,T)\nonumber \\
 &  & +\frac{1}{2h}\int dz_{1}dp_{z2}sin[\frac{1}{\hbar}z_{1}(p_{z}-p_{z2})](\frac{1}{m^{*}(z-\frac{z_{1}}{2})}-\frac{1}{m^{*}(z+\frac{z_{1}}{2})})p_{\parallel}^{2}f_{\downarrow\downarrow}(p_{\parallel},p_{z2},z,T)\nonumber \\
 &  & +\frac{1}{h}\int dz_{1}dp_{z2}sin(\frac{1}{\hbar}z_{1}(p_{z}-p_{z2}))[E_{c}(z-\frac{z_{1}}{2})-E_{c}(z+\frac{z_{1}}{2})]f_{\downarrow\downarrow}(p_{\parallel},p_{z2},z,T)\nonumber \\
 &  & -\frac{\alpha}{\hbar}p_{x}(f_{\uparrow\downarrow}+f_{\downarrow\uparrow})-i\frac{\alpha}{\hbar}p_{y}(f_{\uparrow\downarrow}-f_{\downarrow\uparrow})\nonumber \\
 &  & +\sum_{\sigma=\uparrow,\downarrow}\sum_{\eta=+1,-1}\frac{1}{h^{3}}\int d\mathbf{q}(\delta_{\downarrow\sigma}-f_{\downarrow\sigma}(\mathbf{p}+\mathbf{q},\mathbf{r},T))f_{\sigma\downarrow}({\bf \mathbf{p}},\mathbf{r},T)\nonumber \\
 &  & \times\delta(\frac{E_{\downarrow}({\bf \mathbf{p}}+{\bf \mathbf{q}})+E_{\sigma}({\bf \mathbf{p}}+{\bf \mathbf{q}})}{2}-\frac{E_{\sigma}({\bf \mathbf{p}})+E_{\downarrow}(\mathbf{p})}{2}+\eta\Omega_{\mathbf{q}})M_{\mathbf{q}}^{2}(N_{\mathbf{q}}+\frac{1}{2}-\frac{1}{2}\eta)\nonumber \\
 &  & -\sum_{\sigma=\uparrow,\downarrow}\sum_{\eta=+1,-1}\frac{1}{h^{3}}\int d\mathbf{q}f_{\downarrow\sigma}(\mathbf{{\bf \mathbf{p}}+{\bf \mathbf{q}}},\mathbf{r},T)(\delta_{\sigma\downarrow}-f_{\sigma\downarrow}({\bf \mathbf{p}},\mathbf{r},T))\nonumber \\
 &  & \times\delta(\frac{E_{\downarrow}({\bf \mathbf{p}}+{\bf \mathbf{q}})+E_{\sigma}({\bf \mathbf{p}}+{\bf \mathbf{q}})}{2}-\frac{E_{\sigma}({\bf \mathbf{p}})+E_{\downarrow}(\mathbf{p})}{2}+\eta\Omega_{\mathbf{q}})M_{\mathbf{q}}^{2}(N_{\mathbf{q}}+\frac{1}{2}+\frac{1}{2}\eta).\end{eqnarray}

The equations above can be further simplified by considering a constant
effective mass and ignoring the scattering 

\begin{eqnarray*}
\hbar\frac{\partial f_{\uparrow\uparrow}(p_{\parallel},p_{z},z,t)}{\partial T} & = & -\frac{p_{z}}{m^{*}}\frac{\partial}{\partial z}f_{\uparrow\uparrow}+2\frac{\alpha}{\hbar}p_{x}\textrm{Re}[f_{\uparrow\downarrow}]-2\frac{\alpha}{\hbar}p_{y}\textrm{Im}[f_{\uparrow\downarrow}]\\
 &  & +\frac{1}{h}\int dz_{1}dp_{z2}sin(\frac{1}{\hbar}z_{1}(p_{z}-p_{z2}))[E_{c}(z-\frac{z_{1}}{2})-E_{c}(z+\frac{z_{1}}{2})]f_{\uparrow\uparrow}(p_{\parallel},p_{z2},z,T)\\
\hbar\frac{\partial\textrm{Re}[f_{\uparrow\downarrow}(p_{\parallel},p_{z},z,t)]}{\partial T} & = & -\frac{p_{z}}{m^{*}}\frac{\partial}{\partial z}f_{\uparrow\downarrow}+\frac{\alpha}{\hbar}p_{x}(f_{\downarrow\downarrow}-f_{\uparrow\uparrow})\\
 &  & +\frac{1}{h}\int dz_{1}dp_{z2}sin(\frac{1}{\hbar}z_{1}(p_{z}-p_{z2}))[E_{c}(z-\frac{z_{1}}{2})-E_{c}(z+\frac{z_{1}}{2})]f_{\uparrow\downarrow}(p_{\parallel},p_{z2},z,T)\\
\hbar\frac{\partial\textrm{Im}[f_{\uparrow\downarrow}(p_{\parallel},p_{z},z,t)]}{\partial T} & = & -\frac{p_{z}}{m^{*}}\frac{\partial}{\partial z}f_{\downarrow\uparrow}+\frac{\alpha}{\hbar}p_{y}(f_{\uparrow\uparrow}-f_{\downarrow\downarrow})\\
 &  & +\frac{1}{h}\int dz_{1}dp_{z2}sin(\frac{1}{\hbar}z_{1}(p_{z}-p_{z2}))[E_{c}(z-\frac{z_{1}}{2})-E_{c}(z+\frac{z_{1}}{2})]f_{\downarrow\uparrow}(p_{\parallel},p_{z2},z,T)\\
\hbar\frac{\partial f_{\downarrow\downarrow}(p_{\parallel},p_{z},z,t)}{\partial T} & = & -\frac{p_{z}}{m^{*}}\frac{\partial}{\partial z}f_{\downarrow\downarrow}-2\frac{\alpha}{\hbar}p_{x}\textrm{Re}[f_{\uparrow\downarrow}]+2\frac{\alpha}{\hbar}p_{y}\textrm{Im}[f_{\uparrow\downarrow}]\\
 &  & +\frac{1}{h}\int dz_{1}dp_{z2}sin(\frac{1}{\hbar}z_{1}(p_{z}-p_{z2}))[E_{c}(z-\frac{z_{1}}{2})-E_{c}(z+\frac{z_{1}}{2})]f_{\downarrow\downarrow}(p_{\parallel},p_{z2},z,T)\end{eqnarray*}
where $\textrm{Re}[f_{\uparrow\downarrow}]$ denotes the real part
of $f_{\uparrow\downarrow}$ and $\textrm{Im}[f_{\uparrow\downarrow}]$
denotes the imaginary part of $f_{\uparrow\downarrow}$.

The particle density in each band is written in terms of the diagonal
components of the Wigner function matrix as

\begin{equation}
n_{\sigma}=\frac{1}{h^{3}}\int d\mathbf{p}f_{\sigma\sigma}(\mathbf{p},r),\label{particle}\end{equation}
so that the spin up and spin down particle density are written as 

\begin{eqnarray}
n_{\uparrow} & = & \frac{1}{h^{3}}\int dp_{\parallel}dp_{z}f_{\uparrow\uparrow}(p_{\parallel},p_{z},z),\\
n_{\downarrow} & = & \frac{1}{h^{3}}\int dp_{\parallel}dp_{z}f_{\downarrow\downarrow}(p_{\parallel},p_{z},z).\end{eqnarray}
 The current density can be calculated using the following equation

\begin{equation}
J(r)=\sum_{\sigma,\sigma^{'}}\frac{q}{h^{3}}\int d\mathbf{p}\frac{\partial H_{\sigma\sigma^{'}}}{\partial\mathbf{p}}f_{\sigma^{'}\sigma}(\mathbf{p},r).\label{current}\end{equation}
Therefore the spin up and spin down current components can be written
respectively as 

\begin{eqnarray}
J_{\uparrow}(z) & = & \frac{q}{h^{3}}\int dp_{\parallel}dp_{z}\frac{p_{z}}{m^{*}}f_{\uparrow\uparrow}(p_{\parallel},p_{z},z),\\
J_{\downarrow}(z) & = & \frac{q}{h^{3}}\int dp_{\parallel}dp_{z}\frac{p_{z}}{m^{*}}f_{\downarrow\downarrow}(p_{\parallel},p_{z},z).\end{eqnarray}

We derived the two-spin band Wigner function equations for a Rashba
effect resonant tunneling diode. The effective mass became nonlocal.
This provides the possibility of an accurate study of the relationship
between the position dependent effective mass and the spin splitting
in the resonant tunneling structures since the interface effects on
the effective mass are explicitly included in this model. Then we
simplified the equations using a constant effective mass. This gives
a set of equations that are simpler to solve numerically than the
first set. The current and particle densities were derived. These
equations are going to serve as the starting point for the Wigner
function simulations of zero magnetic field resonant spin tunneling
devices. 

This work was supported by a grant from the Army Research Office's
Defense University Research Initiative on Nanotechnology (DURINT).
We thank Bernard Rosen and Peiji Zhao for useful discussions and suggestions.


\begin{thebibliography}{10}
\bibitem{Winkler2}R. Winkler, Spin-Orbit Coupling Effects in Two-Dimensional Electron
and Hole Systems, (Springer-Verlag, New York, 2003).
\bibitem{Ting}D. Z.-Y. Ting, X. Cartoixa, D. H. Chow, J. S. Moon, D. L. Smith, T.
C. McGill and J. N. Schulman, Proceedings of the IEEE, 91, 741, (2003). 
\bibitem{Winkler}R. Winkler, Phys. Rev. B, 62, 4245, (2000).
\bibitem{Buot}F. A. Buot and K. L. Jensen, Phys. Rev. B, 42, 9429, (1990). 
\bibitem{Erasmos}Erasmos A. de Andrada.e Silva, and Giuseppe C. La Rocca, Phys. Rev.
B, 59, 15583, (1999).
\bibitem{Koga}Takaaki Koga, Junsaku Nitta, Hideaki Takayanagi and Supriyo Datta,
Phys. Rev. Lett., 88, 126601, (2002).
\bibitem{David}David Z. -Y. Ting and Xavier Cartoixa,  Appl. Phys. Lett., 81, 4198,
(2002).
\bibitem{Unlu}M. B. Unlu, B. Rosen, P. Zhao and H. L. Cui, (submitted to Phys. Lett.
A), 2004.
\bibitem{Langreth}D. C. Langreth and J. W. Wilkins, Phys. Rev. B, 6, 3189, (1972).
\bibitem{Koch}R. Binder and S. W. Koch, Prog. Quant. Electr., 19, 307, (1995).
\bibitem{Wu}M. Q. Weng and M. W. Wu, J. Phys: Condens. Matter, 15, 5563, (2003).\end{thebibliography}
\end{document}